# Confirmation of Kramers-Henneberger Atoms

Qi Wei[1], Pingxiao Wang[2], Sabre Kais[3] and Dudley Herschbach[4]

[1]State Key Laboratory of Precision Spectroscopy, East China Normal University, Shanghai 200062, China
[2]Institute of Modern Physics, Department of Nuclear Science and Technology, Fudan University, Shanghai 200433, China.
[3]Department of Chemistry and Physics, Purdue University, West Lafayette, Indiana 47907, USA; Qatar Environment and Energy Research Institute, HBKU, Qatar Foundation, Doha, Qatar.
[4]Department of Chemistry and Chemical Biology, Harvard University, Cambridge, Massachusetts 02138, USA; Institute for Quantum Science and Engineering, Texas A&M University, College Station, Texas 77843, USA

**COMMENT ON** U. Eichmann, T. Nubbemeyer, H. Rottke & W. Sandner *Nature* **461**, 1261–1265 (2009)

In a remarkable experiment, Eichmann *et al.* attained unprecedented acceleration of neutral atoms, up to $10^{15}$ m/s², by strong short-pulse IR laser fields[1]. The driving mechanism was identified as the ponderomotive force on excited electrons bound in Rydberg orbits that survive long enough to enable the atoms to reach the detector. However, the observed velocities lie somewhat above the theoretical prediction. The systematic discrepancy was attributed to "absolute laser intensity uncertainties or a slightly non-Gaussian intensity distribution"[1]. Here, we examine the process by transforming to the Kramers-Henneberger (KH) reference frame[2]. We find that in addition to the ponderomotive potential there exists a smaller but significant term that comes from the binding energy of the KH atom. Including this KH term brings the calculated maximum velocities to a close match with experimental results over the full range of laser pulse durations.

Starting from the Schrödinger equation for a one-electron atom in the velocity gauge, applying the nondipole KH transformation in the nonrelativistic regime[3] recasts the Hamiltonian as:

$$\hat{H} = \frac{\boldsymbol{p}^2}{2} + V(\boldsymbol{r}+\boldsymbol{\alpha}) + \frac{\boldsymbol{A}^2}{2}, \qquad (1)$$

with $\boldsymbol{p}$ the momentum, $\boldsymbol{A}$ the vector potential, and $\boldsymbol{\alpha}(\boldsymbol{t})$ the quiver motion of the electron relative to the laboratory frame of a classical free electron in the laser field. In the KH frame, $V(\boldsymbol{r}+\boldsymbol{\alpha})$ is the potential due to interaction of the electron with the nucleus or to an atomic core, which corresponds to $V(\boldsymbol{r})$ in the lab frame. Since only one electron is excited, the interaction between the loosely bound electron and the atomic core is modeled by[4]:

$$V(\boldsymbol{r}) = -\frac{1}{r}\bigl(1 + e^{-\delta_0 r}\bigr), \qquad (2)$$

comprised of a long range Coulomb potential and a short range Yukawa type potential. The parameter $\delta_0 = 2.13$ and 2.32 (atomic units) for Helium and Neon atoms, respectively[4]. The atom is subject to a linearly polarized laser pulse propagating along the z direction and the corresponding vector potential has the form:

$$\boldsymbol{A}(t) = \frac{E_0(r,t)}{\omega} \sin[\omega(t - z/c)]\hat{x} \qquad (3)$$

The cycle-averaged vector potential term in the Hamiltonian is actually the ponderomotive potential: $\widehat{H}_{PM} = <\boldsymbol{A}^2/2> = |E_0(r,t)|^2/4\omega^2$, which represents the kinetic energy due to oscillation of the electron in the laser field. The other terms in the Hamiltonian, denoted $\widehat{H}_{KH}$, represent the KH atom. Then the force imposed on the atom is:

$$F = -\nabla\langle\widehat{H}\rangle = -\nabla\langle\widehat{H}_{KH}\rangle - \frac{1}{4\omega^2}\nabla|E_0(r,t)|^2 \qquad (4)$$

Only the last term in Eq. (4), the ponderomotive force, $-\nabla\langle\widehat{H}_{PM}\rangle$, was used in Ref. [1] to explain the observed acceleration effect, which is almost entirely in the radial direction, perpendicular to the laser beam. The PM force comes solely from classical dynamics, whereas the KH term arises from quantum mechanics.

To assess the contribution from $-\nabla\langle\widehat{H}_{KH}\rangle$, we assume that the quantum state of the KH atom evolves adiabatically, which holds when the field amplitude varies slowly during the laser pulse[5]. Then the time-dependent dynamics can be converted into a quasistationary Schrödinger equation[6]:

$$\left[\frac{\boldsymbol{p}^2}{2} + V_0(\boldsymbol{r},\alpha_0)\right]\Phi_{KH} = \epsilon_n(\alpha_0)\Phi_{KH} \qquad (5)$$

Here the interaction potential has been "dressed" by averaging the electron quiver motion over a laser oscillation cycle:

$$V_0(\boldsymbol{r},\alpha_0) = \frac{1}{2\pi}\int_0^{2\pi} V[\boldsymbol{r} + \boldsymbol{\alpha}(\xi/\omega)]d\xi, \qquad (6)$$

with $\boldsymbol{\alpha}(t) = \alpha_0 \cos(\omega t)\hat{x}$. The amplitude, $\alpha_0 = \sqrt{I}/\omega^2$, is governed by the laser frequency $\omega$, which is constant, and intensity $I$, for which the spatial and temporal distributions are specified by explicit formulas given in Ref. [1]. The eigenenergies $\epsilon_n$ are functions only of $\alpha_0$ and the corresponding states are termed KH states. Usually, Eq.(5) has been used in treating high-frequency laser fields[5-7], but recent work has shown that for bound KH states, it is a suitable approximation for low-frequency fields[8-10]. Accordingly, we can approximate $\langle\widehat{H}_{KH}\rangle$ by $\epsilon_0(\alpha_0)$, the ground KH state; it is only weakly bound but much more so than higher states. In order to compare with the experimental results, we focus on acceleration along the radial direction, given by:

$$F_{KH} = -\frac{\partial}{\partial r}\langle\widehat{H}_{KH}\rangle = -\frac{\partial\epsilon_0}{\partial\alpha_0} \cdot \frac{\partial\alpha_0}{\partial I} \cdot \frac{\partial I}{\partial r} \qquad (7)$$

Both $\partial\alpha_0/\partial I$ and $\partial I/\partial r$ can be obtained from explicit formulas[1] whereas $\partial \epsilon_n / \partial \alpha_0$ requires numerically solving Eq.(5). It is much larger for the ground state (n = 0) than the higher states in the range of $\alpha_0$ that contributes most to $F_{KH}$, another reason for dealing just with the ground state.

Fig.1a-b shows the KH ground state binding energy $\epsilon_0$ and $\partial \epsilon_0 / \partial \alpha_0$ as functions of the quiver amplitude $\alpha_0$ for He and Ne atoms, compared with results[5] for the H atom. The three curves nearly overlap for $\alpha_0 > 4\ a.u.$ and separate from H only for $\alpha_0 < 4\ a.u.$. As $\alpha_0$ increases, $\epsilon_0$ decreases monotonically, hence the ground state KH atom is a low-field seeker. Fig.1c displays the trajectory of the quiver amplitude during the laser pulse intensity envelope, $f(t) = \exp(-t^2/\tau^2)$, with $\tau_{\text{FWHM}} = 100$ fs. Fig. 1d shows the corresponding PM force and KH force exerted on a He atom located at the focal plane and half beam waist during the laser pulse. Fig.1e-f plots for He and Ne atoms the maximum velocity $V_{max}(z=0)$ imparted to them at the focal plane as a function of laser pulse duration at constant laser intensity. Black dots with error bars are experimental data and dashed lines show theoretical results, including only the PM force, from Ref. [1]. Red lines and dots show our results, including both the PM and KH force contributions. We used the laser parameters specified in Ref. [1], without any fitting or scaling. However, the experimentally measured laser beam waist, $w_0 = 17.5 \pm 1.5$ μm, was appreciably uncertain. Hence, we did calculations for the range of $w_0$, shown by the bars attached to our red points.

As evident in Fig.1d, the overall contribution from the PM force is much stronger than that from the KH force, but their magnitudes differ markedly during the laser pulse. Fig.lc exhibits how the temporal variation corresponds to the quiver amplitude. Thus, $F_{KH}$ is dominant during both the entrance and exit portions of the laser pulse, when $t/\tau > \pm 2$ and $\alpha_0 < 5$ a.u., whereas $F_{PM}$ peaks (and $F_{KH}$ droops to its minimum) at the mid-point of the pulse, when $t = 0$ and $\alpha_0 = 70$ a.u..

The strong radial acceleration observed in Ref. [1] was attributed to excited atomic states, metastable Rydberg states, in which during the laser pulse the excited electron behaves as a quasi-free electron. Thereby the electron charge is accelerated in the laser field (much more than the far heavier nuclear charge), but remains weakly tethered to the atom core, so drags along the center-of-mass. In the KH frame, that tethering scenario pertains even for the ground state. As seen in Fig.1a, It resembles a Rydberg state in having small binding energy and large size. The large electron quiver amplitude attained near peak intensity produces the PM force, while the much smaller quiver domain associated with lower intensity at edges of the laser pulse contributes the KH force.

Our application of the KH frame to the acceleration of laser-dressed atoms provides a simple illustration of its utility. It also serves, together with other evidence, to add to the confirmation of the KH atom as "a physically relevant object in strong IR fields"[8-10].

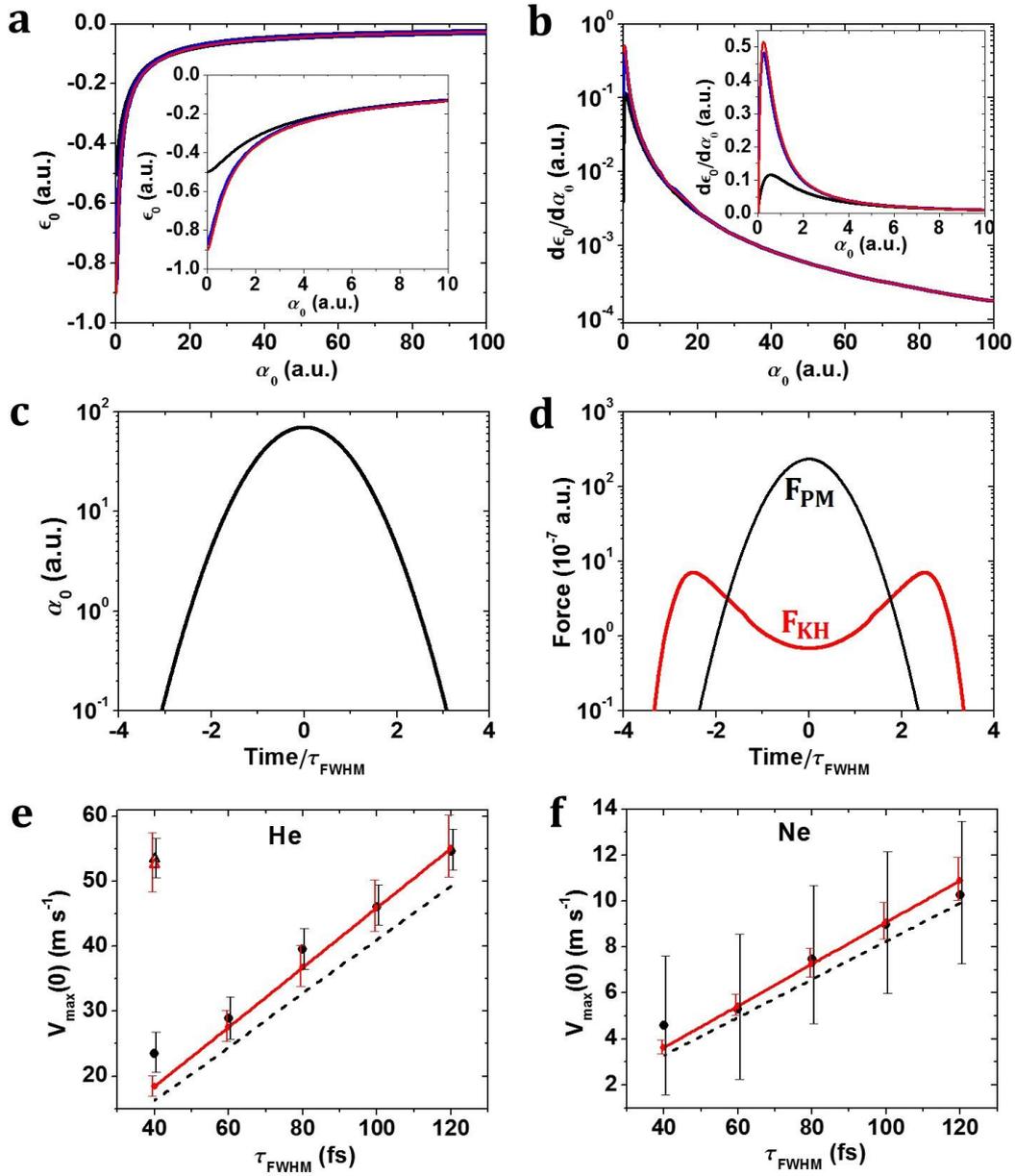

**Figure 1 | Results computed for KH atoms**. **a.** Ground state energy $\epsilon_0$ and **b.** slope $d\epsilon_0/d\alpha_0$ as functions of the quiver amplitude $\alpha_0$. Red is for He; blue for Ne; black for H atom. **c.** Dependence on time during the laser pulse envelope (in units of pulse width, $\tau_{FWHM}$; reaching maximum at t = 0) of the quiver amplitude and **d.** force exerted on a He atom located at the focal plane (z = 0) and half beam size (r = $w_0/2$) during the laser pulse. Black curve is for ponderomotive force, red for KH force. Laser parameters used are: $I_0 = 2.8 \times 10^{15}$ W cm$^{-2}$; $w_0 = 17.5$ μm; $\lambda = 814$ nm; $\tau_{FWHM}$ = 100 fs. **e.** Maximum velocity $V_{max}(0)$ transferred to He and **f.** to Ne at the focal plane as a function of laser pulse duration at constant laser intensity. Black dots with error bars are experimental data and dashed lines show theoretical results from Ref. [1], including only PM force (using $w_0 = 16$ μm). Red lines and dots are from our theoretical calculations, including both PM and KH force contributions (using $w_0 = 17.5$ μm). Bars attached to red dots show range

corresponding to the uncertainty of the laser beam waist (bar top $w_0 = 16$ μm, bottom 19 μm). *Note*: in upperleft corner of **e.** the black triangle shows experimental datum for $I_0 = 8.3 \times 10^{15}$ W cm$^{-2}$ and $\tau_{\text{FWHM}} = 40$ fs, obtained from Fig. 2f of Ref. [1]; red triangle is our corresponding theoretical result.

**Author Contributions** All authors contributed to the theoretical formulation; Q.W. and P.W. performed the calculations; Q.W., S.K. and D.H. prepared the manuscript.

**Author Information** Correspondence should be addressed to Q.W. (qwei@admin.ecnu.edu.cn).